\documentclass{article}

\usepackage{fullpage}

\usepackage{amssymb}                                    
\usepackage{mathrsfs}                    
\usepackage{amsmath}                    
\usepackage{amsfonts}                   
\usepackage{amsthm}                     
\usepackage{makeidx}
\usepackage{graphics}
\usepackage{hyperref}
\usepackage{enumerate}

\usepackage{color}

\usepackage[all]{xy}

\theoremstyle{plain}
\newtheorem{theorem}{Theorem}[section]
\newtheorem{lemma}[theorem]{Lemma}

\newtheorem{proposition}[theorem]{Proposition}
\newtheorem{corollary}[theorem]{Corollary}

\theoremstyle{definition}
\newtheorem{definition}[theorem]{Definition}
\newtheorem{example}[theorem]{Example}

\newtheorem{remark}[theorem]{Remark}

\renewcommand{\(}{\begin{equation}}
\renewcommand{\)}{\end{equation}}
\newcommand{\bea}{\begin{eqnarray}}
\newcommand{\eea}{\end{eqnarray}}
\newcommand{\R}{{\mathbb R}}

\newcommand{\Z}{{\mathbb Z}}
\newcommand{\Q}{{\mathbb Q}}

\makeatletter
\def\@biblabel#1{#1}
\makeatother

\def\proof {\noindent {Proof.}\hspace{7pt}}
\def\endofproof {\hfill{$\Box$}\\ }

\begin{document}

\title{Lie $n$-algebras of BPS charges}
\author{ 
  Hisham Sati\thanks{University of Pittsburgh, Pittsburgh, PA 15260, USA, and New York University, Abu Dhabi, UAE} 
  and 
  Urs Schreiber\thanks{Mathematics Institute of the Academy, {\v Z}itna 25, 115 67 Praha 1, Czech Republic}
}
\maketitle

\begin{abstract}
We uncover higher algebraic structures on Noether currents and BPS charges.
 It is known that equivalence classes of conserved currents form a Lie algebra.
 We show that at least for target space symmetries of higher parameterized WZW-type sigma-models
 this naturally lifts to a Lie $(p+1)$-algebra structure on the Noether
 currents themselves.
  Applied to the Green-Schwarz-type action functionals
  for super $p$-brane sigma-models
  this yields super Lie $(p+1)$-algebra refinements of the
  traditional BPS brane charge extensions of supersymmetry algebras. We discuss this in the generality
  of higher differential geometry, where it applies also to branes with (higher) gauge fields on their worldvolume.
  Applied to the M5-brane sigma-model we recover and properly globalize the M-theory super Lie algebra extension
  of 11-dimensional superisometries by 2-brane and 5-brane charges. Passing beyond the infinitesimal Lie theory we find
  cohomological corrections to these charges in higher analogy to the
  familiar corrections for D-brane charges as they are lifted from ordinary cohomology to twisted K-theory.
  This supports the proposal that M-brane charges live in a  twisted cohomology theory.
\end{abstract}

\tableofcontents



\section{Introduction}

The correspondence between infinitesimal symmetries of a local Lagrangian form and conserved
currents of its induced Euler-Lagrange equations, i.e. Noether's first variational theorem,
 is one of the cornerstones of mathematical field theory.
 If the Lagrangian under consideration is preserved only up to an exact term,
as is the case for sigma-models of Wess-Zumino-Witten-type (WZW),
then the Lie algebra of conserved currents is an extension of the symmetry algebra
(see for instance \cite[section 8]{AzcarragaIzqierdo95} for review from a perspective that will be useful here).
Furthermore,  the extensions arising this way are interesting enough that they acquired
a life of their own in pure mathematics: for the 2-dimensional WZW model they transgress
to the \emph{current algebras} from affine Lie algebra theory (see e.g. \cite{Mi} for a survey).

\medskip
Key examples of
higher-dimensional WZW models are the Green-Schwarz-type sigma
models describing the worldvolume theory of all
super $p$-branes \cite{HenMez,CdAIP,InfinityWZW}.
Their extended current algebras have been argued in \cite{AGT89} to yield
extensions
of superisometry algebras of curved
superspacetimes by brane charges \cite{Townsend95, Hull97}. These constitute
what characterizes BPS states of supergravity and hence of supersymmetric field theories (e.g \cite{CallisterSmith}).

\medskip
However, in traditional constructions one considers a Lie bracket
only on equivalence classes of current forms modulo exact forms, the Dickey bracket \cite{Dickey},
\cite[section 3]{BarnichHenneaux}.
Whenever one is in such a situation where an algebraic structure exists on the homology
of a chain complex, it is natural to ask whether a homotopy-theoretic refinement of
the algebraic structure may exist on the chain complex itself, where algebraic conditions
such as Jacobi identities may hold only up to higher coherent homotopies. For the case of Lie
algebras this means lifting to $L_\infty$-algebras (see \cite{SSSI,LocalObservables} for review and references),
called Lie $(p+1)$-algebras when the underlying chain complex is concentrated in degrees 0 up to $p$.

\medskip
For the case of Lie algebras of conserved Noether currents, a strategy to lift to an $L_\infty$-algebra had been
proposed in \cite{BFLS98,Shnider}. However, the construction there only
applies when all relevant higher charge cohomology groups vanish, resulting in an $L_\infty$-algebra
that is in fact quasi-isomorphic to the plain  Lie algebra of Noether currents. Hence
the general question remained open.

\medskip
What we do here for field theories of higher dimensional and parameterized WZW-type is the following:
\begin{enumerate}
\item We identify the correct higher current algebra for higher WZW-type Lagrangians
as the higher Poisson bracket $L_\infty$-algebra of local observables considered in
\cite{Rogers:2010nw}, for the pre-$(p+1)$-plectic form being the higher WZW curvature form on target spacetime.
In fact we find two different but quasi-isomorphic $L_\infty$-algebras of higher WZW-type Noether currents, following \cite{LocalObservables},
corresponding to the two different incarnations of the Dickey bracket observed in \cite[(3.10)-(3.11)]{BarnichHenneaux}.
\item We show that
the $L_\infty$-Heisenberg-Kostant-Souriau extension of \cite[theorem 3.3.1]{LocalObservables} is the homotopy-theoretic lift
  of the central extension of infinitesimal symmetries via Noether currents by de Rham cohomology classes (``topological currents'')
  as in \cite[p. 203]{ViKr99}.
\item We indicate how the higher stacky Heisenberg-Kostant-Souriau extension of \cite{hgp} generalizes this to
higher gauged WZW models (such as the Green-Schwarz-type sigma-model for the M5-brane) and produces cohomological
corrections to the naive brane charges.
\end{enumerate}

\medskip

\noindent {\bf Acknowledgements}
We thank Domenico Fiorenza for discussion of truncation of $L_\infty$-algebra extensions in section \ref{TruncationToOrdinaryLieAlgebras},
and we thank Jim Stasheff for comments on the text.
U.S. thanks Igor Khavkine for discussion of higher WZW charge algebras in traditional variational calculus;
a joint account dedicated to this aspect is in preparation. U.S. was supported by RVO:67985840.

\newpage

We review some of the background material in the context of local observables in Sec. \ref{LocalObservablesAndHigherPoissonBracketHomotopyLieAlgebras}.
The algebras we discuss in this letter form a hierarchy of the following form (see \cite{SSSI, LocalObservables} for discussion and references).
\begin{itemize}
\item $L_\infty$-algebras $\in L_\infty{\rm Alg}$. These have $k$-ary brackets for all $k \in \mathbb{N}$, $k \geq 2$ on a chain complex;
\item dg-algebras $\in {\rm dgLieAlg} \hookrightarrow L_\infty{\rm Alg}$. These are $L_\infty$-algebras with at most binary brackets,
 hence Lie algebras in the category of chain complexes;
\item Lie $n$-algebras $\in \mathrm{Lie}_n\mathrm{Alg}  \hookrightarrow L_\infty{\rm Alg}$.
These are $L_\infty$-algebras whose underlying chain complex is
concentrated in degree 0 to $(n-1)$, also called $n$-term $L_\infty$-algebras;
\item Lie algebras $\in \mathrm{LieAlg} =  \mathrm{Lie}_1\mathrm{Alg}$. These are the usual Lie algebras, i.e. Lie 1-algebras, viewed as $L_\infty$-algebras or dg-Lie algebras whose underlying chain
complex is concentrated in degree 0.
\end{itemize}

{ This concept of
{\it Lie $n$-algebras} in the homotopy theoretic sense of $n$-term $L_\infty$-algebras, originally due to Stasheff
\cite{LadaStasheff92}, is different from
a vaguely similar definition due to Filippov that has come to be known as {\it $n$-Lie algebras}.
But the Filippov 3-Lie algebras found on the worldvolume of coincident M2-branes
are a special case of metric $L_\infty$-algebras \cite[section 2.5]{RitterSaemann13}.
Notice that (super-)$L_\infty$-algebras play a role in supergravity and
super $p$-brane sigma-models already since \cite{DF}, where they appear in the dual but equivalent guise of their
Chevalley-Eilenberg dg-algebras (``FDA''s), as explained in \cite{SSSI}.}

\medskip
The key motivation for finding an $L_\infty$-algebra of currents is that it retains
more local information about the system at hand. This is in the general spirit of
refining field theories to what are called ``extended'' or ``multi-tiered'' field theories;
we refer the reader to the introductions of \cite{FiorenzaSatiSchreiberCS, LocalObservables}
for more on this general background.
The urge to refine Lie 1-algebras of infinitesimal symmetries to $L_\infty$-algebras becomes more pronounced
as one aims to pass from infinitesimal to finite symmetries. For the example of field theories of
WZW type, the Lagrangian is not a globally defined differential form, but is rather a connection on a higher
bundle (higher gerbe) on field space. Here the only way even to see the full structure of the finite symmetries is
to work in homotopy theory.

\medskip
Once we have the higher algebras characterized as above, we will describe two ways of getting back to
`lower' algebras (in the categorical sense), including ordinary Lie algebras:

\begin{enumerate}
\item {\it Truncation}: In Sec. \ref{TruncationToOrdinaryLieAlgebras}
we consider appropriate truncations of higher algebras down to more classical structures, by controlling the
length the corresponding chain complex.
Given any homotopy-theoretic higher structure, hence a homotopy $n$-type, there its ``truncation'' down to a 0-type,
obtained by quotienting out all symmetries. This operation in general forgets much relevant structure, but
the resulting object is sometimes more easily recognized and compared to more traditional concepts.
We characterize the corresponding current algebra obtained by truncating the higher Poisson
algebras.

\item {\it Transgression}: We explain in Sec. \ref{Transgression} how the full $L_\infty$-algebra of conserved
currents admits maps to all its transgressions, hence to the results of evaluating
conserved currents on cycles of any dimension. For example, the current Lie 2-algebra
of the 2-dimensional WZW model transgresses to the ``affine'' centrally extended loop Lie algebra of currents
by integrating currents over a circle. The problem of obtaining the 1-algebraic structure of loop spaces
from a 2-algebraic (stacky, gerby) structure on the original base spaces by transgression had been highlighted
prominently in \cite[p. 249]{Brylinski}.

\end{enumerate}

Then in Sec. \ref{BraneChargesAndSupergravityBPSStates}  we discuss the application of the above
construction to super $p$-brane sigma-models.
\begin{itemize}
\item  Application  to the Green-Schwarz-type action functionals
  for super $p$-brane sigma-models
yields super Lie $(p+1)$-algebra refinements of the
  traditional BPS brane charge central extensions of supersymmetry algebras.
  We discuss this in the generality
  of higher geometry where it applies also to branes with (higher) gauge fields on their worldvolume.

\item   Specifically, application to the M5-brane sigma-model recovers the M-theory super Lie algebra extension
  of 11-dimensional superisometries by 2-brane and 5-brane charges.


\item  Passing beyond the infinitesimal Lie theory we find
  cohomological corrections to the above charges in higher analogy to the
  familiar corrections for D-brane charges as they are lifted from ordinary cohomology to twisted K-theory.
  This supports the proposal that M-brane charges live in a  twisted cohomology theory \cite{Duality, KSpin, Sati10}
  and is also supported by the companion analysis in \cite{FSS15}.
\end{itemize}

\section{Poisson bracket Lie $n$-Algebras}
\label{LocalObservablesAndHigherPoissonBracketHomotopyLieAlgebras}

We start by briefly reviewing the pertinent concepts and results from \cite{LocalObservables} in a way that makes
manifest their relation to conserved currents.
In order to see the origin of the higher symmetry in current algebras fully transparently, it is
useful to adopt the following general perspective.

\medskip
Traditionally a local Lagrangian for a $(p+1)$-dimensional field theory is thought of as a suitable differential $(p+1)$-form
$\mathbf{L}$ on the jet bundle of the field bundle of the theory.
(If a reference volume form $\mathbf{vol}$
on spacetime/worldvolume is chosen, then $\mathbf{L}$ is proportional to that form, $\mathbf{L} = L \mathbf{vol}$,
and the coefficient function $L$ is what in many textbooks is regarded as the Lagrangian.)
However, it is well known, though perhaps not widely appreciated, that for some key field theories of interest
(locally variational theories \cite{FPW11}) it
is not in fact globally true that the Lagrangian is a differential $(p+1)$-form. Rather in general it is
on locally so, but is globally a $(p+1)$-form connection on a ``$(p+1)$-bundle'' or ``$p$-gerbe''.
This is notably the case for field theories of
Wess-Zumino-Witten type; see \cite{InfinityWZW} for review of this fact and for pointers to the literature.

\medskip
Even if $\mathbf{L}$ happens to be given by a globally defined $(p+1)$-form, it is still useful to regard this
as the $(p+1)$-connection form on a trivial $(p+1)$-bundle over field space, because it is this perspective that
makes immediately clear that and how there are higher symmetries-of-symmetries.

\medskip
In order to capture this state of affairs mathematically, it is useful to consider a simple object with a
pretentious name: the universal smooth moduli stack of $(p+1)$-connections,
which we denote by $\mathbf{B}^{p+1}U(1)_{\mathrm{conn}}$.
The reader may find exposition and introduction for such objects in the context of field theory in \cite{FiorenzaSatiSchreiberCS},
but this object is easily described and understood already by its defining universal property.
That is,
being a {\it smooth universal moduli stack for higher connections}
 means precisely  that it is
a generalized smooth space of sorts, with the following characterizing properties:
for $X$ any smooth manifold, then
\begin{enumerate}
  \item smooth functions $\nabla \colon X \longrightarrow \mathbf{B}^{p+1}U(1)_{\mathrm{conn}}$
   correspond equivalently to $(p+1)$-form connections on $X$;
   \item
     smooth homotopies between such functions
     $
       
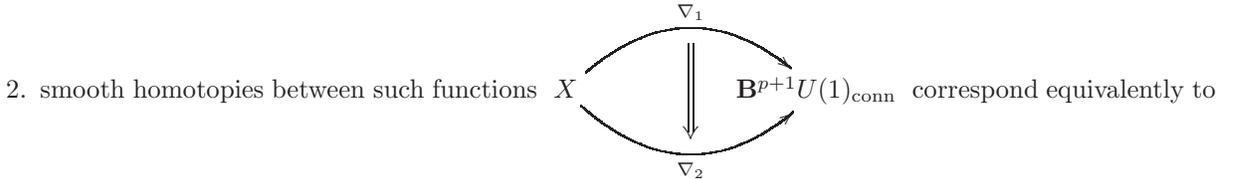
\xymatrix{
          X
          \ar@/^2pc/[rr]^{\nabla_1}_{\ }="s"
          \ar@/_2pc/[rr]_{\nabla_2}^{\ }="t"
          &&
          \mathbf{B}^{p+1}U(1)_{\mathrm{conn}}
          \ar@{=>} "s"; "t"
       }
     $
     correspond equivalently to gauge transformations between two such connections $\nabla_1$ and $\nabla_2$;
   \item
    smooth homotopies-of-homotopies correspond to gauge-of-gauge transformations,
    and so forth, up to $p$-fold homotopies-of-homotopies.
\end{enumerate}

When the connections $\nabla_i$ are given by globally defined differential $(p+1)$-forms $\theta_i$, then
a gauge transformation between these is nothing but a $p$-form $\Delta$ such that
$\theta_2 = \theta_1 + \mathbf{d} \Delta$, where $\mathbf{d}$ denotes the de Rham differential.
A gauge-of-gauge transformation is accordingly $(p-1)$-form, etc. up to order-$p$ gauge-of-gauge transformation
which are (not 0-forms but) $U(1)$-valued functions.

\begin{remark}
{\bf (i)}
One immediate advantage of ``modulating'' connections and differential forms by maps into an object
such as $\mathbf{B}^{p+1}U(1)_{\mathrm{conn}}$ is that it makes their ``contravariance'' manifest while
keeping track of higher gauge transformations. Namely, given a $(p+1)$-form connection $\nabla$ on $X$ and
given a map $\phi : Y \longrightarrow X$ between spaces, then the pullback connection $\phi^\ast \nabla$
is simply the one modulated by the composite map
$$
  \phi^\ast \nabla :
  Y
  \stackrel{\phi}{\longrightarrow}
  X
  \stackrel{\nabla}{\longrightarrow}
  \mathbf{B}^{p+1}U(1)_{\mathrm{conn}}
  \,.
$$

\noindent {\bf (ii)} So suppose then that $X$ is some space of fields of a field theory, typically the jet bundle to the field bundle.
Then it is clear what a {\it symmetry} of such $\nabla$ should be: namely a diffeomorphism
$$
  \xymatrix{
    X \ar[rr]^\phi_\simeq && X
  }
$$
of the field space, such that $\nabla$ is preserved by this up to a specified gauge transformation
$\mathbf{\eta} : \phi^\ast \nabla \stackrel{\simeq}{\longrightarrow} \nabla$.

\noindent {\bf (ii)}
In the diagrammatic language that we have set up, this just means that a symmetry is a pair $(\phi,\mathbf{\eta})$
labeling a diagram of the form
$$
   \raisebox{20pt}{
  \xymatrix{
    X \ar[rr]^\phi_{\ }="s" \ar[dr]_{\nabla}^>>>{\ }="t" && X\;. \ar[dl]^{\nabla}
    \\
    & \mathbf{B}^{p+1} U(1)_{\mathrm{conn}}
    \ar@{=>}^\simeq_{\mathbf{\eta}} "s"; "t"
  }}
  \,
$$
\end{remark}
This is equivalent to saying that if $\nabla$ happens to be given by a globally defined $(p+1)$-form $\theta$,
then the homotopy is given by a $p$-form $\eta$ such that
$$
  \phi^\ast \theta - \theta = \mathbf{d}\eta
  \,.
$$
Moreover, if we have a smooth 1-parameter flow $t \mapsto (\phi(t), \mathbf{\Delta}(t))$ of such symmetries,
then differentiating this relation with respect to $t$ yields the expression of an infinitesimal symmetry of $\nabla$:
$$
  \mathcal{L}_v \theta = \mathbf{d}\Delta_v\;,
$$
where $v$ is the vector field of the flow $t\mapsto \phi(t)$, with $\mathcal{L}_v$ its
Lie differentiation, and where $\Delta_v = \frac{d}{dt}\eta$.
Notice that if one thinks of $\theta = \mathbf{L}$ as being the Lagrangian form
of a field theory, then this is precisely the expression for a ``weak'' symmetry of the Lagrangian, i.e.
a transformation that leaves the Lagrangian invariant up to addition of an exact term $\mathbf{d}\Delta_v$.

\medskip
It is useful to equivalently rewrite this as follows: let $\omega = F_{\nabla}$ denote the curvature $(p+2)$-form of the
$(p+1)$-form connection. If the connection is given by a globally defined differential $(p+1)$-form $\theta$,
then this is simply $\omega = \mathbf{d}\theta$. Cartan's formula then gives that
the contraction is exact
$$
  \iota_v \omega = -\mathbf{d}J_v
$$
with
$$
 J_v := \iota_v \theta - \Delta_v
$$
a kind of Legendre transform of the weakening term.

\medskip

If we do think of $\mathbf{L} = \theta$ as being a WZW-type local Lagrangian, then the expression
$J_v = \iota_v \theta - \Delta_v$ is the conserved Noether charge induced by the
symmetry $v$. This is the interpretation that we are interested in here, as it is what
connects to the conceptualization of BPS charges from Noether currents of higher WZW terms
in \cite{AGT89}.

(More generally. when $\mathbf{L}$ is not just a WZW term, then one has to refine the discussion,
take $X$ above to be a jet bundle and incorporate the horizontal/vertical decomposition of the variational bicomplex
of forms on that bundle. This further generality will be discussed elsewhere, here
we focus on topological terms of higher WZW type.)

For $p = 0$  the above relations are well-known from symplectic geometry, for the case that $\omega$ is a
(pre-)symplectic form. Since this is the correct analogy, we also say that a closed $(p+2)$-form
$\omega$ is a \emph{multisymplectic} $(p+2)$-form or a \emph{$(p+1)$-plectic form} \cite{Rogers:2010nw}.
In view of this it is sensible to agree on the following terminology, modeled directly on the
traditional case given by $p = 0$.

\begin{definition}
  \label{dPlecticHamiltonianVectorFieldInIntroduction}
{\bf (i)}  A closed $(p+2)$-form $\omega$ on a manifold $X$ is also called a \emph{pre-$(p+1)$-plectic form},
  and the pair $(X,\omega)$ a \emph{pre-$(p+1)$-plectic manifold}.

 \noindent {\bf (ii)} Given such, then
  a vector field $v$ such that $\mathcal{L}_v \omega = 0$ called a \emph{pre-$(p+1)$-plectic vector field}.

  \noindent {\bf (iii)} If, moreover, there exists a $p$-form $J$ with $\iota_v \omega = - \mathbf{d}J$
  then $v$ is called a \emph{Hamiltonian vector field} and $J$ is called a \emph{Hamiltonian form}
  for $v$. The pair $(v,J)$ is then a \emph{Hamiltonian pair}.

  \noindent {\bf (iv)} If $\omega$ is non-degenerate in that
  the kernel of the contraction $\iota_{(-)}\omega : \mathrm{Vect}(X) \to \Omega^{p+1}(X)$ vanishes, then $v$ is
  uniquely determined by its current $J$; in this case we call $\omega$ \emph{$(p+1)$-plectic}
  (instead of just pre-$(p+1)$-plectic).
  \end{definition}

  This is the case that is often restricted to in the literature,
  but the whole theory goes through in the pre-plectic case using such Hamiltonian pairs instead of just Hamiltonian forms.

  \begin{definition}
  \label{SpaceOfHamiltonianPairs}
  {\bf (i)} We write
  $$
    \Omega^{p}_{\omega}(X)
	:=
	\left\{
	  \left(
	    v, J
	  \right)
	  \in \mathrm{Vect}(X)\oplus \Omega^{p}(X)
	  \;|\;
	  \iota_v \omega = - \mathbf{d}J
	\right\}
  $$
  for the linear space of the above pairs and speak of \emph{local Hamiltonian observables}.

 \noindent {\bf (ii)} We write
  $$
    \mathrm{Vec}_{\mathrm{Ham}}(X,\omega) \hookrightarrow \mathrm{Vect}_{\mathrm{Symp}}(X,\omega) \hookrightarrow \mathrm{Vect}(X)
  $$
  for the subspaces of {\it symplectic} and {\it Hamiltonian vector fields}.
  \end{definition}

  Notice that both inclusions above are sub-Lie algebras under the canonical Lie bracket of vector fields.
When regarding $\theta = \mathbf{L}$ as a local Lagrangian,  one then finds that
the contraction $\iota_{\cdots}\omega$ vanishes on tangents to field trajectories which solve the
corresponding equations of motion. Consequently, $J_v$ is an on-shell conserved current, induced
by the given symmetry. This is a special case of the general $(p+1)$-plectic Noether theorem
\cite{dcct}.

\medskip
Next, symmetries as above form a group in an evident way, where the composition operation
is the pasting of diagrams
\(
  \xymatrix{
    X \ar[rr]^-{\phi_1}_>{\ }="s1" \ar[drr]_{\nabla}^{\ }="t1" &&
    X \ar[d]|{\nabla}^{\ }="t2" \ar[rr]^-{\phi_2}_{\ }="s2" && X \ar[dll]^\nabla
    \\
    && \mathbf{B}^{p+1}U(1)_{\mathrm{conn}}
    \ar@{=>}^{\eta_1} "s1"; "t1"
    \ar@{=>}^{\eta_2} "s2"; "t2"
  }
  \,.
  \label{CompositonDiagram}
\)
Working out the Lie bracket of this Lie group one finds (as a special case of what is generally given by Def. \ref{PoissondgAlgebra}
below) that the bracket on the
these Hamiltonian pairs $(v,\Delta_v)$ (from Def. \ref{dPlecticHamiltonianVectorFieldInIntroduction}) is given by
$$
  [(v_1,\Delta_{v_1}), (v_2,\Delta_{v_2}) ]
  =
  ([v_1,v_2], \;  \mathcal{L}_{v_1} \Delta_{v_2} - \mathcal{L}_{v_2} \Delta_{v_1} )
  \,.
$$
\begin{remark}
 \label{AGTFormulaRecovered}
In the case that $\nabla$ is given by a globally defined form $\theta$
and we think of this as being a local Lagrangian as before, so that the $\Delta$s are its conserved
Noether currents, then after passing to de Rham cohomology classes, the above bracket may be identified with what is
known as the \emph{Dickey bracket} on Noether currents \cite{Dickey} \cite[(3.10)]{BarnichHenneaux}.
We generalize this to the case where $\omega$ is not assumed to be globally exact
below in Prop. \ref{BracketUnderSplitting}.

Moreover, when a fixed potential $\Delta_{[v_1,v_2]}$ entering the definition of $J_{[v_1,v_2]}$ is picked
from its equivalence class, then in terms of this the above bracket becomes
$$
  [(v_1,\Delta_{v_1}), (v_2,\Delta_{v_2}) ]
  =
  ([v_1, v_2], \Delta_{[v_1,v_2]}) + (0,  \mathcal{L}_{v_1} \Delta_{v_2} - \mathcal{L}_{v_2} \Delta_{v_1} - \Delta_{[v_1,v_2]})\;.
$$
This may be identified with the bracket on Noether currents as displayed in \cite[(13), (14)]{AGT89} for the case
of super $p$-brane sigma-models. We turn to this example below in Sec. \ref{BraneChargesAndSupergravityBPSStates}.

\end{remark}

\medskip
Therefore, we have found traditional current algebras from a diagrammatic calculus of symmetries of
higher connections.

\begin{remark}
This highlights two points which seem not to have been explicitly addressed in traditional literature:

\begin{enumerate}
\item {\it Higher groups:}
When $p > 0$, the group of symmetries is a higher group (a higher group stack) of
higher order symmetries-of-symmetries. Namely the homotopies $\eta$ in the diagram
\eqref{CompositonDiagram} may themselves have
homotopies-of-homotopies between them
$$
   \raisebox{20pt}{
  \xymatrix{
    X \ar[rr]^\phi_{\ }="s" \ar[dr]_{\nabla}^>>>{\ }="t" && X \ar[dl]^{\nabla}
    \\
    & \mathbf{B}^{p+1} U(1)_{\mathrm{conn}}
    \ar@{}|{\stackrel{}{\Rightarrow}} "s"; "t"
    \ar@/_1pc/@{=>} "s"; "t"
    \ar@/^1pc/@{=>} "s"; "t"
    %
  }}
  \,
$$
corresponding to gauge-of-gauge transformations.
Hence after Lie differentiation then, on top of the Lie bracket of conserved
currents above, there are
higher order gauge transformations between these currents. These
may be most directly understood from the
fact that the choice of $\Delta_v$ above is clearly only unique up to addition of exact terms,
whose potentials in turn are themselves only unique up to exact terms, and so forth.
As a result, we find not just a Lie algebra, but a dg-Lie algebra of currents, whose
differential is the de Rham differential acting on higher order current forms.

\item {\it Refined cohomology and higher forms:}
In full generality, the above discussion needs to be performed
not just for globally defined $\theta$, but for higher prequantizations $\overline{\theta}$
which are given by {\v C}ech-Deligne cocycles with curvature $(d+1)$-form $\omega$.
This allows for the inclusion of geometric data via differential cohomology.
\end{enumerate}
\end{remark}

Incorporating these, the resulting dg-Lie algebra is the one given
in \cite[Def./Prop. 4.2.1]{LocalObservables}:

\medskip
   Let $X$ be a smooth manifold, $\omega \in \Omega^{p+2}_{\mathrm{cl}}(X)$
   a closed differential $(p+2)$-form, with $(X,\omega)$ regarded as a pre-$(p+1)$-plectic manifold.
   Let $\overline{\theta}$ be a higher pre-quantization of $\omega$ given by a {\v C}ech-Deligne cocycle
   with respect to a cover $\mathcal{U}$ of $X$. Write $\mathrm{Tot}^\bullet(\mathcal{U},\Omega^\bullet)$
   for the {\v C}ech-de Rham complex of the cover and notice that the Lie derivatives of $\overline{\theta}$
   are naturally regarded as taking values in this complex.  Then we have:

\begin{definition}
   \label{PoissondgAlgebra}
 The
   \emph{Poisson bracket dg-Lie algebra}
   $$
     \mathfrak{Pois}_{\mathrm{dg}}(X,\overline{\theta})
     \in
     \mathrm{dgLieAlg}
     \hookrightarrow
     L_\infty\mathrm{Alg}
   $$
   is the dg-Lie algebra whose underlying chain complex has components
\bea
     \mathfrak{Pois}_{\mathrm{dg}}(X,\overline{\theta})^0
    & :=&
     \left\{
       (v,\overline{\Delta}) \in \mathrm{Vect}(X) \oplus \mathrm{Tot}^{p}(\mathcal{U},\Omega^\bullet) \vert
       \mathcal{L}_v \overline{\theta} = \mathbf{d}_{\mathrm{Tot}}\overline{\Delta}
     \right\}\;,
 \nonumber\\
     \mathfrak{Pois}_{\mathrm{dg}}(X,\overline{\theta})^{i \geq 1}
    & :=&
     \mathrm{Tot}^{p-i}(\mathcal{U},\Omega^\bullet)\;,
   \nonumber
   \eea
   with differential $\mathbf{d}_{\mathrm{Tot}}$, and whose
   non-vanishing Lie brackets are
\bea
     [(v_1, \overline{\Delta}_1), (v_2,\overline{\Delta}_2)]
     &=&
     \left([v_1,v_2], \; \mathcal{L}_{v_1} \overline{\Delta}_2 - \mathcal{L}_{v_2} \overline{\Delta}_1 \right)\;,
\nonumber\\
     \left[(v, \overline{\Delta}), \overline{\eta}\right]
     &=& - [\overline{\eta}, (v,\overline{\Delta})] = \mathcal{L}_v \overline{\eta}\;.
     \nonumber
   \eea
\end{definition}

It turns out that there is a very different looking but equivalent incarnation of this
$L_\infty$-algebra, originally considered in
\cite{Rogers:2010nw}:

\begin{definition}[Higher Poisson bracket of local observables]
  \label{PoissonLienAlgebra}
 \label{ham-infty}
  Given a pre-$(p+1)$-plectic manifold $(X,\omega)$,

  \noindent {\bf (i)} we say that the truncated de Rham complex ending in the Hamiltonian pairs
  of Def. \ref{SpaceOfHamiltonianPairs}
  is the \emph{complex of local observables} of $(X,\omega)$, denoted
  $$
    \Omega^\bullet_\omega(X)
	:=
    \left(
	  C^\infty(X)
	  \stackrel{\mathbf{d}}{\longrightarrow}
	  \Omega^1(X)
	  \stackrel{\mathbf{d}}{\longrightarrow}
	  \cdots
	  \stackrel{\mathbf{d}}{\longrightarrow}
	  \Omega^{n-2}(X)
	  \stackrel{(0,\mathbf{d})}{\longrightarrow}
	  \Omega^{p}_\omega(X)
	\right)
	\,,
  $$
  with $\Omega^{p}_\omega(X)$ in degree zero.

  \noindent {\bf (ii)} The \emph{binary higher Poisson bracket} on local Hamiltonian observables
  is the linear map
  $$
    \{-,-\} \;:\; \Omega^{p}_\omega(X) \otimes \Omega^{p}_\omega(X) \longrightarrow
	 \Omega^{p}_\omega(X)
  $$
  given by the formula
  \begin{equation}\label{eq.Lie-bracket}
    \left[
	  \left(v_1, J_1\right),
	  \left(v_2, J_1\right)	
	\right]
	:=
	\left[
	  \left(
	    \left[v_1, v_2\right],
		\,
		\iota_{v_1 \wedge v_2}\omega
	  \right)
	\right]
	\,;
  \end{equation}

\noindent {\bf (iii)}  and for $k \geq 3$
  the \emph{$k$-ary higher Poisson bracket} is the
  linear map
  $$
    \{-,\cdots, -\}
	\;:\;
    \left( \Omega^{p}_\omega(X)\right)^{\otimes^k}
	\longrightarrow
	\Omega^{p+2-k}(X)
  $$
  given by the formula
  $$
    \left[
	  \left(v_1, J_1\right),\,
	  \cdots,
	  \left(v_k, J_k\right)
	\right]
	:=
	(-1)^{\lfloor \frac{k-1}{2}\rfloor}
	\iota_{v_1 \wedge \cdots \wedge v_k}
	\omega
	\,.
  $$
\noindent {\bf (iv)}  The chain complex of local observables equipped with these linear maps
  for all $k$ will be called
  the \emph{higher Poisson bracket $L_\infty$ algebra} of
  $(X,\omega)$, denoted
  $$
    \mathfrak{Pois}_\infty(X,\omega)
	:=
	\left(
	  \Omega^\bullet_\omega(X),\mathbf{d},
	  \left\{-,-\right\}, \left\{-,-,-\right\}, \cdots
	\right)
	\,.
  $$
\end{definition}

\begin{remark}
  \label{EquivalenceOfTheTwoModelsForHigherPois}
From \cite[Theorem 4.2.2]{LocalObservables}, there is an equivalence of $L_\infty$-algebras
  $$
    \mathfrak{Pois}_\infty(X,\omega)
    \stackrel{\simeq}{\longrightarrow}
    \mathfrak{Pois}_{\mathrm{dg}}(X,\omega)
  $$
  between those of Def. \ref{PoissondgAlgebra} and Def. \ref{PoissonLienAlgebra}.
\end{remark}

We would like to gain some further understanding of these $L_\infty$-algebras.
The idea is that starting with $\R$, we can view it as a chain complex by placing it
in various degrees with the remaining degrees being zero. Shifting the degrees
up and down result in stack analogs of the classifying functor ${\bf B}$. When the resulting
stack is flat, we denote this with a musical $\flat$.

\begin{definition}
For $X$ a smooth manifold, denote
\begin{enumerate}
\item by $\mathbf{H}(X,\flat\mathbf{B}^{p}\mathbb{R})$ the abelian
 Lie $(p+1)$-algebra given by the chain complex
 $$
\Omega^0(X)\!\xrightarrow{\mathbf{d}}\Omega^1(X)\!\xrightarrow{\mathbf{d}}\cdots \!\xrightarrow{\mathbf{d}}\Omega^{p}(X)\;,
$$
with $\Omega^p(X)$ in degree 0;
\item by $\mathbf{B}\mathbf{H}(X,\flat\mathbf{B}^{p}\mathbb{R})$
the abelian Lie $(p+1)$-algebra given by the chain complex, including one more term,
$$
\Omega^0(X)\!\xrightarrow{\mathbf{d}}\Omega^1(X)\!\xrightarrow{\mathbf{d}}\cdots \!\xrightarrow{\mathbf{d}}\Omega^{p}(X)\!\xrightarrow{\mathbf{d}}\mathbf{d}\Omega^{p}(X)\;,
$$
with $\mathbf{d}\Omega^{p}(X)$ in degree zero.
\end{enumerate}
\label{ResolvedDeloopedShiftedTruncatedDeRhamComplex}
\end{definition}

We start with relating the latter to Hamiltonian vector fields.
\begin{remark}
  \label{prop.just-above}
From {\cite[Proposition 3.2.3]{LocalObservables}}
there is the following result.
For $(X,\omega)$ a pre-$(p+1)$-plectic manifold, the multilinear maps
\begin{eqnarray*}
  \omega_{[1]} &:& v\mapsto -\iota_v\omega;
\nonumber\\
  \omega_{[2]} &:& v_1\wedge v_2\mapsto \iota_{v_1\wedge v_2}\omega;
  \nonumber\\
     \vdots\\
     \nonumber
  \omega_{[p+2]} &:& v_1\wedge v_2\wedge\cdots  v_{p+2}\mapsto -(-1)^{\binom{p+2}{2}}
\iota_{v_1\wedge v_2\wedge \cdots \wedge v_{p+2}}\omega
\nonumber
\end{eqnarray*}
define an $L_\infty$-morphism
$
  \omega_{[\bullet]}
  :
\mathrm{Vect}_{\mathrm{Ham}}(X)\xrightarrow{} \mathbf{B} \mathbf{H}(X,\flat\mathbf{B}^{p}\mathbb{R})\,.
$
This is the higher Kirillov-Kostant-Souriau $L_\infty$-alge\-bra $(p+2)$-cocycle on the Lie algebra of
Hamiltonian vector fields with
values in the abelian $(p+2)$-algebra $\mathbf{B} \mathbf{H}(X,\flat\mathbf{B}^{p}\mathbb{R})$ associated to the pre-$(p+1)$-plectic manifold $(X,\omega)$.
\end{remark}

The $L_\infty$-algebras in Remark \ref{EquivalenceOfTheTwoModelsForHigherPois}
are characterized as follows.

\begin{remark}
\label{TheFullExtension}
From {\cite[Theorem 3.3.1]{LocalObservables}}, the Poisson bracket Lie $(p+1)$-algebra $\mathfrak{Pois}_\infty(X,\omega)$ is
 the extension of the Lie algebra of Hamiltonian vector fields, Def. \ref{dPlecticHamiltonianVectorFieldInIntroduction},
  by the abelian Lie $(p+1)$-algebra $\mathbf{H}(X,\flat\mathbf{B}^{p}\mathbb{R})$
  induced by the Kirillov-Kostant-Souriau $L_\infty$-alge\-bra $(p+2)$-cocycle  $\omega_{[\bullet]}$.
  Namely,   there is a homotopy fiber sequence of $L_\infty$-algebras of the form
    $$
    \raisebox{20pt}{
    \xymatrix{
      \mathbf{H}(X,\flat\mathbf{B}^{p}\mathbb{R} )
      \ar[r]
      &
      \mathfrak{Pois}_\infty(X,\omega)
      \ar[d]
      \\
      & \mathrm{Vect}_{\mathrm{Ham}}(X,\omega)
      \ar[r]^-{\omega[\bullet]}
      &
      \mathbf{B}\mathbf{H}(X,\flat\mathbf{B}^{p}\mathbb{R} )\;.
    }
    }
  $$
  Here the vertical homomorphism comes from the projection $\Omega^{p}_\omega(X)\to \mathrm{Vect}_{\mathrm{Ham}}(X)$
(from Hamiltonian pairs to Hamiltonian vector fields, Def. \ref{SpaceOfHamiltonianPairs})
which induces a surjective linear morphism of $L_\infty$-algebras.
\end{remark}

\section{Truncation to ordinary Lie algebras}
\label{TruncationToOrdinaryLieAlgebras}

We would like to consider appropriate truncations down to more classical structures. This is governed by controlling the
`size' of the corresponding chain complex.
Given any homotopy-theoretic higher structure, hence a homotopy $n$-type, there its ``truncation'' down to a 0-type,
obtained by quotienting out all symmetries. This operation in general forgets much relevant structure, but
the resulting object is sometimes more easily recognized and compared to more traditional concepts.

\medskip
For the simple special case of chain complexes of vector spaces in non-negative degree (in the homological
degree conventions), 0-truncation is the functor
$$
  \tau_{\leq 0}
  :
  \mathrm{Ch}_\bullet
  \longrightarrow
  \mathrm{Vect}
$$
which sends a chain complex $(\cdots V_2 \stackrel{\partial_1}{\to}V_1 \stackrel{\partial_0}{\to} V_0)$
to its degree-0 homology group
$$
  \tau_{\leq 0} V_\bullet
  =
  V_0/ \mathrm{im}(\partial_0)
  =
  H_0(V_\bullet)
  \,.
$$

We now discuss this 0-truncation for the Lie $(p+1)$-algebras of conserved currents discussed in the previous
section, Sec. \ref{LocalObservablesAndHigherPoissonBracketHomotopyLieAlgebras}.

\begin{lemma}\label{lemma.truncation}
The 0-truncation functor $\tau{\leq 0}$ on chain complexes induces a functor
\[
\tau_{\leq 0}\colon L_\infty\text{-}\mathrm{alg}_{\geq 0}\to \mathrm{Lie}
\]
where $L_\infty\text{-}\mathrm{alg}_{\geq 0}$ denotes the category of $L_\infty$-algebras concentrated in nonnegative degrees, with $L_\infty$-morphisms as morphisms, and $\mathrm{Lie}$ denotes the category of Lie algebras with Lie algebra morphisms. More explicitly, $\tau_{\leq 0}$ maps an $L_\infty$-algebra $(\mathfrak{g},\mathbf{d},
	  \left\{-,-\right\}, \left\{-,-,-\right\}, \cdots
	)$ concentrated in nonpositive degrees to the Lie algebra $(H_0(\mathfrak{g}), \left\{-,-\right\})$ and an $L_\infty$-morphism $f\colon \mathfrak{g}\to \mathfrak{h}$ to the Lie algebra morphism $H_0(f_1)\colon H_0(\mathfrak{g})\to H_0(\mathfrak{h})$, where $f_1$ is the linear component of $f$. Moreover, the natural morphism of chain complexes $\mathfrak{g}\to \tau_{\leq 0}\mathfrak{g}$ is a natural linear and surjective $L_\infty$-morphism
	\(
	(\mathfrak{g},\mathbf{d},
	  \left\{-,-\right\}, \left\{-,-,-\right\}, \cdots
	)\to (H_0(\mathfrak{g}), \left\{-,-\right\})\;.
	\label{Surjective}
	\)
	\end{lemma}
\proof
Since the chain complex $\mathfrak{g}$ is concentrated in nonnegative degrees, the chain complex $\tau_{\leq 0}\mathfrak{g}$ consists of the vector space $H_0(\mathfrak{g})$ concentrated in degree zero. All the statements in the lemma then follow straightforwardly from this.
\endofproof

\begin{remark}
Naturality of the linear $L_\infty$-morphism
\eqref{Surjective} means that for any $L_\infty$-morphism $f\colon \mathfrak{g}\to \mathfrak{h}$ between $L_\infty$-algebras concentrated in nonnegative degree we have a commutative diagram of $L_\infty$-morphisms
	\[
	  \raisebox{20pt}{
    \xymatrix{
     \mathfrak{g}\ar[d]
      \ar[r]^{f}
      &
      \mathfrak{h}
      \ar[d]
      \\
       H_0(\mathfrak{g})
      \ar[r]^-{ H_0({f_1})}
      &
       H_0(\mathfrak{h})\;.
    }
    }
	\]
\end{remark}

We now look at the 0-truncation of the corresponding Poisson algebras.

\begin{definition}
  \label{The0TrunctionOfThePoissonBracket}
  Let $(X,\omega)$ be a pre-$(p+1)$-plectic manifold.
  Write
  $$
    \mathfrak{Pois}(X,\omega) := \tau_{\leq 0}\mathfrak{Pois}_\infty(X,\omega)
  $$
  for the 0-truncation of
  $\mathfrak{Pois}_\infty(X,\omega)\simeq \mathfrak{Pois}_{\mathrm{dg}}(X,\omega)$.
  \end{definition}
  Note that using  Def. \ref{PoissondgAlgebra}, Def. \ref{PoissonLienAlgebra}, Remark \ref{EquivalenceOfTheTwoModelsForHigherPois},
  and via Lemma \ref{lemma.truncation}, this is the quotient $\mathfrak{Pois}_\infty(X,\omega) \to H_0\mathfrak{Pois}_\infty(X,\omega)$
  by exact current $p$-forms. We  now characterize this 0-truncation.

\begin{proposition}
  \label{0TruncationOfHigherPoisExtension}
  Let $(X,\omega)$ be a pre-$(p+1)$-plectic manifold.
  Then the Lie 1-algebra of currents, Def. \ref{The0TrunctionOfThePoissonBracket},
  is a central extension of the Hamiltonian vector fields by the abelian Lie algebra
  $H^{p}_{\mathrm{dR}}(X)$ of de Rham $p$-forms. In other words, writing $\mathfrak{Pois}(X,\omega)$ for
   $H_0\mathfrak{Pois}_\infty(X,\omega)$, there is a short exact sequence of Lie algebras
  $$
    0 \to H^{p}_{\mathrm{dR}}(X) \longrightarrow  \mathfrak{Pois}(X,\omega)
     \longrightarrow
     \mathrm{Vect}_{\mathrm{Ham}}(X,\omega)
     \to 0
     \,,
  $$
  and $H^{p}_{\mathrm{dR}}(X)$ is central in $ \mathfrak{Pois}(X,\omega)$.
\end{proposition}
\proof
From the short exact sequence of chain complexes given by Prop. \ref{TheFullExtension}
\[
0\to \mathbf{H}(X,\flat\mathbf{B}^{p}\mathbb{R} )\to \mathfrak{Pois}_\infty(X,\omega) \to \mathrm{Vect}_{\mathrm{Ham}}(X,\omega)
     \to 0\;,
\]
we get the long exact sequence in homology
\[
 0 \to H^{p}_{\mathrm{dR}}(X)\to \mathfrak{Pois}(X,\omega) \to \mathrm{Vect}_{\mathrm{Ham}}(X,\omega)
     \to 0
\]
and, by Lemma \ref{lemma.truncation}, this is a short exact sequence of Lie algebras. The fact
that $H^{p}_{\mathrm{dR}}(X)$ is central in $ \mathfrak{Pois}(X,\omega)$ is immediate from Equation (\ref{eq.Lie-bracket}).
\endofproof

\begin{remark}
  For the special case when $X$ is the jet bundle of a field bundle, when the
  de Rham differentials appearing everywhere are constrained to be the corresponding horizontal
  differentials, and under the assumption that the cohomology of this horizontal de Rham complex
  is concentrated in degree 0, then another $L_\infty$-stucture on the truncated de Rham complex
  appearing in Def. \ref{PoissonLienAlgebra} has been constructed in \cite{BFLS98}.
  Inspection of the construction around Theorem 7 there shows that this $L_\infty$-algebra is
  $L_\infty$-equivalent to its 0-truncation $\mathfrak{Pois}(X,\omega)$ which under these
  assumptions is the Dickey algebra \cite{Dickey}; see the comments above in
  Sec. \ref{LocalObservablesAndHigherPoissonBracketHomotopyLieAlgebras}. Hence for all purposes
  of homotopy theory the algebra in \cite{BFLS98} \emph{is} this 0-truncation.
\end{remark}

In order to further compare Proposition \ref{0TruncationOfHigherPoisExtension} to existing literature, we consider now a choice of
Hamiltonian current forms for each symmetry generator.
The following statement is for def. \ref{PoissonLienAlgebra} what remark \ref{AGTFormulaRecovered} was for def. \ref{PoissondgAlgebra}.

\begin{proposition}
\label{BracketUnderSplitting}
Under a choice of linear splitting $J\colon v \mapsto (v,J_v )$ of the natural projection $\Omega_\omega^{p}(X)\to \mathrm{Vect}_{\mathrm{Ham}}(X,\omega)$ from Def. \ref{dPlecticHamiltonianVectorFieldInIntroduction},
the Lie bracket on de Rham cohomology classes of currents
in $\mathfrak{Pois}(X,\omega)$, Proposition \ref{0TruncationOfHigherPoisExtension}, is isomorphic to
\[
\{v+[\alpha],w+[\beta]\}=[v,w]+[J_{[v,w]}-\frac{1}{2}(\mathcal{L}_vJ_w-\mathcal{L}_wJ_v)]
\,.
\]
When moreover $\omega$ is globally exact via $\mathbf{d}\theta=\omega$, then in terms of the currents $\Delta_v := i_v\theta-J_v$
this reduces to
\[
\{v+[\alpha],w+[\beta]\}=[v,w]+[\mathcal{L}_v\Delta_w-\mathcal{L}_w\Delta_v-\Delta_{[v,w]}]
\]
as in remark \ref{AGTFormulaRecovered}.
\end{proposition}
\proof
The splitting induces a linear isomorphism of the form
\begin{align*}
\mathrm{Vect}_{\mathrm{Ham}}(X,\omega)\oplus H^{p}_{\mathrm{dR}}(X)&\xrightarrow{\sim}
\mathfrak{Pois}(X,\omega)\\
v+[\alpha]&\mapsto (v,J_v-\alpha)
\end{align*}
 and with this the Lie bracket on $\mathrm{Vect}_{\mathrm{Ham}}(X,\omega)\oplus H^{p}_{\mathrm{dR}}(X)$ reads
\[
\{v+[\alpha],w+[\beta]\}=[v,w]+[J_{[v,w]}-\iota_{v \wedge w}\omega]\;.
\]
Since $\iota_v\omega=-\mathbf{d}J_v$, one has
\begin{equation*}
\iota_{v \wedge w}\omega=-\iota_w\mathbf{d}J_v=-\mathcal{L}_wJ_v+\mathbf{d}\iota_wJ_v
\end{equation*}
and
\begin{equation*}
\iota_{v \wedge w}\omega=-\iota_{w \wedge v}\omega=\iota_v\mathbf{d}J_w=\mathcal{L}_vJ_w-\mathbf{d}\iota_vJ_w\;.
\end{equation*}
Hence, combining, we can write
\[
\iota_{v \wedge w}\omega=\frac{1}{2}(\mathcal{L}_vJ_w-\mathcal{L}_wJ_v)-\frac{1}{2}\mathbf{d}(\iota_vJ_w-\iota_wJ_v)
\]
and so
\[
\{v+[\alpha],w+[\beta]\}=[v,w]+[J_{[v,w]}-\frac{1}{2}(\mathcal{L}_vJ_w-\mathcal{L}_wJ_v)]\;.
\]
If, moreover, a global potential $\theta$ for the pre-$n$-plectic form $\omega$ is given, i.e., if one has a
 $p$-form $\theta$ with $\mathbf{d}\theta=\omega$ then, writing $J_v=i_v\theta-\Delta_v$, one finds
$$
\begin{aligned}
  J_{[v,w]}-\frac{1}{2}(\mathcal{L}_v J_w-\mathcal{L}_w J_v)
  & =
  -
  \Delta_{[v,w]}
  +
  \iota_{[v,w]}\theta
  - \frac{1}{2}(
    \mathcal{L}_v \iota_w \theta
    -
    \mathcal{L}_w \iota_v \theta
  )
  +
  \frac{1}{2}(
    \mathcal{L}_v \Delta_w
    -
    \mathcal{L}_w \Delta_v
  )
  \\
  & =
  -
  \Delta_{[v,w]}
  +
  \underset{= 0}{\underbrace{
  \iota_{[v,w]}\theta
  - \frac{1}{2}(
    \iota_{[v,w]} \theta
    -
    \iota_{[w,v]} \theta
  )
  }}
  - \frac{1}{2}(
    \iota_w \mathcal{L}_v \theta
    -
    \iota_v \mathcal{L}_w \theta
  )
  +
  \frac{1}{2}(
    \mathcal{L}_v \Delta_w
    -
    \mathcal{L}_w \Delta_v
  )
  \\
  & =
  -
  \Delta_{[v,w]}
  - \frac{1}{2}(
    \iota_w \mathbf{d} \Delta_v
    -
    \iota_v \mathbf{d} \Delta_w
  )
  +
  \frac{1}{2}(
    \mathcal{L}_v \Delta_w
    -
    \mathcal{L}_w \Delta_v
  )
  \\
  & = -\Delta_{[v,w]}
   + \mathcal{L}_v\Delta_w-\mathcal{L}_w\Delta_v
   +\text{$\mathbf{d}$-exact terms}
   \;.
\end{aligned}
$$
Therefore, finally, we get the desired expression
\[
\{v+[\alpha],w+[\beta]\}=[v,w]+[\mathcal{L}_v\Delta_w-\mathcal{L}_w\Delta_v-\Delta_{[v,w]}]\;.
\]
\endofproof

\section{Transgression to ordinary Lie algebras}
\label{Transgression}

We will now consider another process of getting ordinary Lie algebras starting from
our higher algebras.
While the 0-truncation which we considered in Section
\ref{TruncationToOrdinaryLieAlgebras} just quotients out and forgets
higher gauge transformations to obtain a plain Lie 1-algebra of currents,
there is another way to turn a higher current Lie $(p+1)$-algebra into a Lie $k$-algebra
for $k \leq p$ that retains more information: mathematically this is
\emph{transgression}, while in terms of physics this corresponds to
the natural operation of integrating local currents over submanifolds of some codimension.
A notable
instance is to do so over Cauchy surfaces, in order to turn them into ordinary observables
(i.e. functions on field configurations).

\medskip
The conception of higher groups of currents as indicated
in Section \ref{LocalObservablesAndHigherPoissonBracketHomotopyLieAlgebras}, via automorpisms sliced over
a higher differential moduli stack, lends itself to a clear picture of the
process of transgression of higher currents as follows.

\medskip
First observe that transgression of plain differential forms has the
following neat sheaf-theoretic formulation. Write $\mathbf{\Omega}^{p+1}$ for the
sheaf of smooth differential $(p+1)$-forms (on the site of all smooth manifolds).
Then, by the Yoneda lemma, a smooth differential $(p+1)$-form $\omega \in \Omega^{p+1}(X)$
on some smooth manifold $X$ is equivalently a morphism in the category of sheaves
of the form
$$
  \omega : X \longrightarrow \mathbf{\Omega}^{p+1}
  \,.
$$
Now, given a closed oriented smooth manifold $\Sigma$ of dimension $k \leq p+1$, write
$[\Sigma,\mathbf{\Omega}^{p+1}]$ for the mapping space, which is the sheaf that sends
a test manifold to the set of sheaf morphisms $U \times \Sigma \to \mathbf{\Omega}^{p+1}$,
and hence to the set of smooth differential $(p+1)$-forms on $U \times \Sigma$.
It is immediate then that ordinary fiber integration of differential forms
along the projections $\Sigma\times U \to U$, for all possible $U$, is embodied
in a single morphism of sheaves of the form
$$
  \int_\Sigma \;:\; [\Sigma,\mathbf{\Omega}^{p+1}] \longrightarrow \mathbf{\Omega}^{p+1-k}
  \,.
$$
Now, forming the mapping sheaf out of $\Sigma$ is a functorial process, and hence we
obtain from the above ingredients the following

\begin{proposition}
There a composite morphism of the form
$$
  \int_\Sigma [\Sigma,\omega]
  \;:\;
  [\Sigma, X]
  \stackrel{[\Sigma,\omega]}{\xrightarrow{\hspace*{1cm}}}
  [\Sigma,\mathbf{\Omega}^{p+1}]
  \stackrel{\int_\Sigma}{\longrightarrow}
  \mathbf{\Omega}^{p+1-k}\;,
$$
representing a differential $(p+1-k)$-form on the smooth mapping space $[\Sigma,X]$.
\end{proposition}

The point now is that, unwinding all the definitions, one finds
\begin{corollary}
The
differential $(p+1-k)$-form $\int_\Sigma [\Sigma,\omega]$ is the transgression
of $\omega$ to the above mapping space, namely is the form
$\int_\Sigma \mathrm{ev}_\Sigma^\ast \omega$, for
$$
  \mathrm{ev}_\Sigma
  \;:\;
  [\Sigma,X]\times \Sigma
  \longrightarrow
  X
$$
the evaluation map.
\end{corollary}

While this is the traditional formulation of
transgression, the above re-formulation via mapping sheaves has the
advantage that it neatly generalizes from differential $n$-forms $\omega$
to their pre-quantization by $(p+1)$-connections $\nabla$,
that we considered in Sec. \ref{LocalObservablesAndHigherPoissonBracketHomotopyLieAlgebras}.

\medskip
Recall that  fiber integration of differential forms, viewed as an operation on all
possible $\Sigma\times U$ as $U$ varies, can be encapsulated in a single morphism of
sheaves as above. Similarly, fiber integration of $n$-connections, hence fiber integration
of cocycles in ordinary differential cohomology of degree $(p+2)$, is equivalently
encoded \cite{FiorenzaSatiSchreiberIV} in a morphism of smooth higher stacks of the form
$$
  \int_\Sigma
  :
  [\Sigma, \mathbf{B}^{p+1}U(1)_{\mathrm{conn}}]
  \longrightarrow
  \mathbf{B}^{p+1-k}U(1)_{\mathrm{conn}}
  \,.
$$
With this in hand we immediately obtain

\begin{proposition}
{\bf (i)} There is a natural homomorphism from the smooth $(p+1)$-group of conserved
currents as in Sec. \ref{LocalObservablesAndHigherPoissonBracketHomotopyLieAlgebras},
to $(p+1-k)$-group of their \emph{transgression} to the mapping space out of $\Sigma$.

\noindent {\bf (ii)}
This is done simply by applying the $\infty$-functor $[\Sigma,-]$ to the defining slice diagrams and then
postcomposing the slicing with $\int_X$, as indicated here:
$$
  \left(
   \raisebox{20pt}{
  \xymatrix{
    X \ar[rr]^\phi_{\ }="s" \ar[dr]_{\nabla}^>>>{\ }="t" && X \ar[dl]^{\nabla}
    \\
    & \mathbf{B}^{p+1} U(1)_{\mathrm{conn}}
    \ar@{=>}^\simeq_{\mathbf{\eta}} "s"; "t"
  }}
  \right)
  \;\;\;
  \mapsto
  \;\;\;
  \left(
   \raisebox{38pt}{
  \xymatrix{
    [\Sigma,X] \ar[rr]^{[\Sigma,\phi]}_{\ }="s" \ar[dr]_{[\Sigma,\nabla]}^>>>{\ }="t" && [\Sigma,X] \ar[dl]^{[\Sigma,\nabla]}
    \\
    & [\Sigma,\mathbf{B}^{p+1} U(1)_{\mathrm{conn}}]
    \ar[d]^{\int_\Sigma}
    \\
    & \mathbf{B}^{p+1-k}U(1)_{\mathrm{conn}}
    \ar@{=>}^\simeq_{[\Sigma,\eta]} "s"; "t"
  }}
  \right)
  \,.
$$
\end{proposition}

\begin{remark}
The above operation now transgresses not only the $(p+1)$-connection $\nabla$, which
for us is the Lagrangian of the given field theory, but transgresses also all the
currents which are embodied in the equivalences labeled $\eta$. Specifically,
if $\nabla$ happens to be given by a globally defined differential $(p+1)$-form
$\theta = \mathbf{L}$ then $\eta$ is, infinitesimally, given by a current
$p$-form $J$, and the above operation takes that to its transgression form
$\int_\Sigma \mathrm{ev}^\ast J$.

\begin{enumerate}
\item When $k = p$, then this transgressed form is a 0-form, hence a function, and has
the interpretation of the actual observable induced by the current $J$.

\item Generally, for $k = p$ then on the right all higher homotopies vanish, due to the composition with $\int_\Sigma$
which lands in the 1-stack $\mathbf{B}U(1)_{\mathrm{conn}}$,
and hence in this case the truncation map factors through the 0-truncation of the group on the left,
which we just discussed above. This way the 0-truncation of the current Lie $(p+1)$-algebra
maps into the generalized affine Lie algebra of currents, but the latter may in general be
bigger.
\end{enumerate}
\end{remark}

\begin{example}
The archetypical example of this process is given by the case of the 3d WZW model on
a compact Lie group $G$. In this case $\nabla$ above is the WZW gerbe
$\mathbf{L}_{\mathrm{WZW}} : G \longrightarrow \mathbf{B}^2 U(1)_{\mathrm{conn}}$ on the group, and its
transgression $\int_{S^1} [S^1, \mathbf{L}_{\mathrm{WZW}}] : [S^1, G] \to \mathbf{B}U(1)_{\mathrm{conn}}$
is a bundle with connection on the loop group. By the results of
\cite{hgp} the group of currents induced by this, as discussed above, is the
``quantomorphism group'' of this bundle regarded as a prequantum line bundle in the
sense of Souriau. It is a famous fact \cite{PrSe88} (see \cite[around Prop. 42]{Segal}),
that this is the Kac-Moody central extension of the loop group of $G$, whose Lie algebra
is the corresponding affine Lie algebra of $G$. This is ``the'' current algebra as seen in
most of the mathematics literature.
\end{example}

We see here that this Kac-Moody loop group is nothing but the transgression of the
smooth 2-group of currents of the WZW model, which, by \cite[2.6.1]{hgp},
is what is called the String 2-group of $G$, and that the
affine current Lie algebra is nothing but the transgression of the corresponding Lie
2-algebra, which is the $\mathfrak{string}$ Lie 2-algebra \cite{BaezRogers}.

\begin{remark}
Two variants of transgression are of relevance in physics:
\begin{enumerate}
\item When the target space is
a  product  $X \times \Sigma$,   there is a canonical morphism
$X \longrightarrow [\Sigma, X \times \Sigma]$. Transgression followed by precomposition with this
inclusion
$$
  (X \times \Sigma \stackrel{\nabla}{\longrightarrow} \mathbf{B}^{p+1}U(1)_{\mathrm{conn}})
    \;\;\mapsto \;\;
  (X \to [\Sigma, X \times \Sigma]
  \stackrel{\int_\Sigma [\Sigma,\nabla]}{\xrightarrow{\hspace*{1.2cm}}}
    \mathbf{B}^{p+1-\mathrm{dim}(\Sigma)}U(1)_{\mathrm{conn}} )
$$
is \emph{double dimensional reduction}.
\item Moreover, when $\Sigma$ is equipped with a base point $s_0 : \ast \to \Sigma$, then there is also the
induced restriction operation
$$
  (X \times \Sigma \stackrel{\nabla}{\longrightarrow} \mathbf{B}^{p+1}U(1)_{\mathrm{conn}})
    \;\; \mapsto \;\;
  (X \to [\Sigma, X \times \Sigma]\stackrel{|_{s_0}}{\xrightarrow{\hspace*{5mm}}} X \times \Sigma \stackrel{\nabla}{\longrightarrow}
  \mathbf{B}^{p+1}U(1)_{\mathrm{conn}})
  \,.
$$
\end{enumerate}
\end{remark}

\begin{example}
The product case occurs frequently in the M-theory circle bundle over 10-dimensional spacetime $X$.
The combination of double dimensional reduction with restriction is, for instance, what takes the 11-dimensional
supergravity C-field to the B-field and Ramonf-Ramond (RR) form in 10 dimensions. The above process
allows us to do so at the level of stacks, where all the higher geometric and gauge structures are carried
along. A stack perspective on this can be found in \cite{FSS-E8}.
\end{example}

\section{Application to brane charges}
\label{BraneChargesAndSupergravityBPSStates}

We now discuss how the above constructions lead to interesting applications in (supersymmetric)
physical models.
We note that all the previous discussion generalizes directly to the situation of
superalgebra and supergeometry, where all manifolds are generalized to supermanifolds,
differential forms to super-differential forms, and  where Lie $n$-algebras are generalized to super Lie $n$-algebras
(see \cite{InfinityWZW} for details and pointers to the literature).
This means that we may apply the previous results to supermanifolds $X$ as they appear in higher supergravity
theory, where they carry certain special super $(p+2)$-forms which serve as the curvatures of WZW terms for
the super $p$-brane sigma-models of Green-Schwarz type, with target space $X$.

\medskip
We consider this in detail. For $N$ a real spin representation of $\mathrm{Spin}(d-1,1)$, let
$\mathbb{R}^{d-1,1|N}$ be the corresponding super-Minkowski spacetime, regarded as
a super translation Lie group. The canonical left-invariant 1-form on $\mathbb{R}^{d-1,1\vert N}$
is traditionally denoted $E = (E^a, \psi^\alpha)_{a,\alpha}$, where $a$ ranges over linear basis elements
for $\mathbb{R}^d$ and $\alpha$ runs over a basis for the linear representation space $N$.
To make precise statements, let then $(d,N,p)$ an item in the old brane scan, i.e. let $p \in \mathbb{N}$
be such that
$$
  \omega_{\mathrm{WZW}}
   :=
  \overline{\psi} \wedge \Gamma_{a_1 \cdots a_p} \psi \wedge E^{a_1} \wedge \cdots \wedge E^{a_p}
  \in \Omega^{p+2}_{\mathrm{cl}}(\mathbb{R}^{d-1,1|N})
$$
is a super Lie algebra cocycle of the given form (see \cite{InfinityWZW} for review and details).
This is then the WZW-curvature term for the Green-Schwarz sigma model for super $p$-branes
propagating on $\mathbb{R}^{d-1,1\vert N}$.

\medskip
For instance for $d = 10$, $p = 1$ and $N = (2,0) = \mathbf{16} + \mathbf{16}$ there is such
a cocycle, and it constitutes the curvature 3-form of the $\kappa$-symmetry WZW-term of the type IIB
superstring on Minkowski background.
More generally, let $X$ be a more general supermanifold locally modeled on $\mathbb{R}^{d-1,1\vert N}$
and equipped with a super-vielbein field (super soldering form) $E$ and further relevant fields,
including a super $(p+2)$-form $\omega$. Then the equations of motion of $d$-dimensional
$N$-supersymmetric supergravity constrain the vielbein to be such that the fermionic part
$\omega^X_{\mathrm{WZW}}$ of $\omega^X$ is the pullback of $\omega_{\mathrm{WZW}}$
via $E$ \cite{BST86, BST87}.

\medskip
Following terminology familiar from the theory of $G_2$-manifolds, we may express this by saying
that $\omega^X_{\mathrm{WZW}}$ is a \emph{definite form}, definine on $\omega_{\mathrm{WZW}}$.
We may regard the pair $(X,\omega^X_{\mathrm{WZW}})$ as a pre-$(p+1)$-plectic supermanifold
according to Def. \ref{dPlecticHamiltonianVectorFieldInIntroduction}. As such it induces
the higher Poisson bracket super Lie $(p+1)$-algebra $\mathfrak{Pois}(X,\omega^X_{\mathrm{WZW}})$ of
Def. \ref{PoissondgAlgebra}, Def. \ref{PoissonLienAlgebra}, and
Remark \ref{EquivalenceOfTheTwoModelsForHigherPois}.

\medskip
The fact that the equations of motion of supergravity force $\omega^X_{\mathrm{WZW}}$ to be
definite on $\omega_{\mathrm{WZW}}$ means that
isometries of $X$ preserve $\omega^X_{\mathrm{WZW}}$, hence that their `super' vector fields
(the Killing vectors and Killing spinors) are $(p+1)$-plectic vector fields in the sense
of Def. \ref{dPlecticHamiltonianVectorFieldInIntroduction}. It is not guaranteed
that every isometry with respect to the given vielbein preserves $\theta$ up to gauge transformation; those
that do are precisely the \emph{Hamiltonian isometries} in the sense of Def. \ref{dPlecticHamiltonianVectorFieldInIntroduction}.
Write
$$
  \mathfrak{Isom}_{\mathrm{Ham}}(X,\omega^X_{\mathrm{WZW}})
  \hookrightarrow
  \mathrm{Vect}_{\mathrm{Ham}}(X,\omega^X_{\mathrm{WZW}})
$$
for the inclusion of these into all Hamiltonian vector fields

\begin{definition}
Write $\mathfrak{BPS}(X,\omega^X_{\mathrm{WZW}})$ for the restriction of the current Lie $(p+1)$-algebra
$\mathfrak{Pois}(X,\omega^X_{\mathrm{WZW}})$
(Def. \ref{PoissondgAlgebra}, Def.\ref{PoissonLienAlgebra}, Remark \ref{EquivalenceOfTheTwoModelsForHigherPois})
of $\omega^X_{\mathrm{WZW}}$ to isometries, i.e. for the super $L_\infty$-algebra in the homotopy pullback
diagram
$$
  \raisebox{20pt}{
  \xymatrix{
    \mathfrak{BPS}(X,\omega^X_{\mathrm{WZW}}) \ar[r] \ar[d] & \mathfrak{Pois}(X,\omega^X_{\mathrm{WZW}}) \ar[d]
    \\
    \mathfrak{Isom}(X,\omega^X_{\mathrm{WZW}})
    \ar[r]
    &
    \mathrm{Vect}_{\mathrm{Ham}}(X,\omega^X_{\mathrm{WZW}})\;.
  }
  }
$$
\end{definition}

\medskip
The following reproduces and generalizes the result in \cite[p.8]{AGT89}.

\begin{proposition}
\label{ThePBraneExtensionAlgebra}
The 0-truncation to a super-Lie algebra $\tau_0 \mathfrak{BPS}(X,\omega)$ is the central extension of the
supersymmetry algebra of $X$ by charges of $p$-branes wrapping non-trivial cycles.
\end{proposition}
\proof
This follows via Remark \ref{BracketUnderSplitting}
by Proposition \ref{0TruncationOfHigherPoisExtension},
which gives an extension
$$
  H^{p}_{dR}(X) \to \tau_0 \mathfrak{BPS}(X,\omega^X_{\mathrm{WZW}})\to
  \mathfrak{Isom}(X,\omega^X_{\mathrm{WZW}})\;,
$$
classified by $\omega^X_{\mathrm{WZW}}(-,-)$. The elements in $H^{p}_{\mathrm{dR}}(X)$ are the $p$-brane charges.
\endofproof

Hence we have the following picture
$$
\left\{
   \raisebox{20pt}{
  \xymatrix{
    X \ar[rr]^\phi_{\ }="s" \ar[dr]_{\mathbf{L}_{\mathrm{WZW}}}^>>>{\ }="t" && X
     \ar[dl]^{\mathbf{L}_{\mathrm{WZW}}}
    \\
    & \mathbf{B}^{p+1} U(1)_{\mathrm{conn}}
    \ar@{=>}^\simeq_{\mathbf{\eta}} "s"; "t"
  }}\;,
  \hspace{5mm}
  \omega_{\mathrm{WZW}}^X = F_{\mathbf{L}_{\mathrm{WZW}}}
  \right\}
  \,
  \stackrel{\mathrm{Lie}}{\mapsto}
  \raisebox{42pt}{
  \xymatrix{
  H_{\rm dR}^p(X)
  \ar[d]
  \\
  \tau_0 {\mathfrak{Poiss}} (X, \omega_{\mathrm{WZW}}^X)
  \ar[d]
  \\
  \mathfrak{Isom}(X,\omega_{\mathrm{WZW}}^X)
   }
  }
$$

%

These are the supersymmetry extensions induced by a single brane species, as considered originally in
\cite{AGT89}. But the ``type II algebra'' and the ``M-theory algebra'' \cite{Townsend95, Hull97} are supposed to arise
from considering not just strings and membranes, but also the branes on which these may end, namely the D-branes
and the M5-brane, respectively.
Notably the M-theory supersymmetry algebra is given on the fermionic generators in traditional
local component notation as \cite{Townsend95, ACDvP03}
\(
\{Q_\alpha, Q_\beta \}=(C\Gamma^M)_{\alpha \beta} P_M + (C\Gamma_{MN})_{\alpha \beta} Z_2^{MN} +
(C\Gamma_{MNPQR})_{\alpha \beta} Z_5^{MNPQR}\;,
\label{susy algebra}
\)
where $C$ is the charge conjugation matrix and $\Gamma$ are the Dirac matrices.
The three terms on the right hand side
correspond to the graviton momentum 1-form charge, the membrane 2-form charge and the fivebrane 5-form charge.
Notice again that this traditional expression applies only locally, on patches of spacetime diffeomorphic to super-Minkowski
spacetime.

Now the proper global analysis of \cite[p.8]{AGT89} and of prop. \ref{ThePBraneExtensionAlgebra} only ever produces
extensions by charges of a single brane species with no (higher) gauge fields on its worldvolume.
But in \cite{InfinityWZW, FSS15, WZWterms}
we explained that the WZW-type sigma models for super $p_2$-branes with (higher) gauge fields on the their
worldvolume and on which super $p_1$-branes may end, are globally defined not on target superspacetime $X$ itself,
but on the total space $\widetilde {\hat X}$ of a super $p_1$-stack extension $\widetilde {\hat X} \to X$
of superspacetime, which itself is a differential refinement of the $p_1$-gerbe $\hat X$ that underlies the
WZW term of the $p_1$-branes.
Moreover, in \cite{hgp} we showed that the
higher Heisenberg-Kostant-Souriau extensions of remark \ref{TheFullExtension} generalizes to such higher stacky base spaces.
Schematically this follows now by the following picture, the full details are in \cite{dcct, WZWterms}:
$$
  \left\{
   \raisebox{60pt}{
  \xymatrix{
  \hat{X} \ar[d] \ar@/_2.9pc/[dddr]_{\mathbf{L}_{\mathrm{WZW}}^{p_2}} \ar[rr]^\simeq && \hat{X} \ar[d] \ar@/^2.9pc/[dddl]^{\mathbf{L}_{\mathrm{WZW}}^{p_2}}
  \\
    X \ar[rr]^\simeq_{\ }="s" \ar[dr]|{\mathbf{L}_{\mathrm{WZW}}^{p_1}}^>>>{\ }="t" && X
     \ar[dl]|{\mathbf{L}^{p_1}_{\mathrm{WZW}}}
    \\
    & \mathbf{B}^{p_1+1} U(1)_{\mathrm{conn}} \ar[d]
    %
    \\
    &  \mathbf{B}^{p_2+1} U(1)_{\mathrm{conn}}
  }}
  \right\}
  \stackrel{\simeq}{\mapsto}
  \raisebox{48pt}{
  \xymatrix{
    H^{p_2}(\widetilde{\hat X})
    \ar[d]
    \\
    \tau_0 \mathfrak{Poiss}(\widetilde{\hat X},\mathbf{L}_{\mathrm{WZW}}^{p_2})
    \ar[d]
    \\
    \mathfrak{Isom}(\widetilde {\hat X}, \mathbf{L}_{\mathrm{WZW}}^{p_2} )
  }}
  \,.
 $$
 (All 2-cells on the left are filled by homotopies of higher stacks. We suppress
 them notationally just for convenience and readability.)

Here the differential refinement $\widetilde {\hat X}$ of $\hat X$ is what makes a sigma-model field
$\Sigma_{p_2} \longrightarrow \widetilde{\hat X}$ be a pair consisting of an ordinary map to
target spacetime $\Sigma_{p_2} \longrightarrow X$ together with a twisted $p_1$-form gauge field
on $\Sigma_6$. This is the global model for super $p$-branes with tensor multiplet higher gauge fields
on their worldvolume.

While this is necessary for the full picture, the isometry group and the
cohomology of $\widetilde{\hat X}$
is hard to compute. There is however a canonical forgetful map $\widetilde{\hat X} \to {\hat X}$
to the geometric realization of this differential stack (regarded itself as a locally constant stack, see \cite{dcct}
for details). By combining results of \cite{dcct} and \cite{Pavlov}, one finds that in the case at hand
$\hat X$ is the homotopy type of the $K(\mathbb{Z},p_1+1)$-fiber bundle over spacetime $X$ that is classified by the
integral class of the background field of the $p_1$-brane.
We now compute the cohomology of that geometric realization.
In a full discussion one will have to pull the result of the following
computation back along the above map to $\widetilde{\hat X}$,
and kernel and cokernel of this pullback map potentially yield yet further corrections to the brane charges.

\medskip

We specialize to the case $p_1 = 2$ and $p_2 = 5$ corresponding to M5-branes propagating in an M2-brane condensate
\cite{InfinityWZW}.
First, consider the cohomology of the fiber $K(\Z, 3)$. At the
integral level, this is known but has a complicated structure. We will instead
consider the corresponding rational cohomology, which is much more accessible.
Indeed, it directly follows from the Hurewicz theorem and the universal coefficient
theorem that
$$
H^k(K(\Z, 3); \Q) =\left\{
\begin{array}{ll}
\Q & {\rm for}~k=0, 3\;, \\
0 & \mbox{otherwise}\;.
\end{array}
\right.
$$


\medskip
Given the homotopy fiber sequence
$$
\xymatrix{
K(\Z, 3) \ar[r] & \hat{X} \ar[d]
\\
& X
}
$$
the cohomology Serre spectral sequence takes the form
$$E_2^{p,q}=H^p(X); H^q(K(\Z, 3)) \Rightarrow H^{p+q}(\hat{X})\;.$$
From the cohomology of the fiber determined above, we see that
$q$ has to be either 0 or 3 in order to contribute.
The relevant differential $d_r: E_2^{p,q} \to E_2^{p+r, q-r+1}$ is then
$d_4$, which raises the cohomology degree by 4.

\medskip
Hence the brane charge extension of the 11-dimensional superisometries is, rationally, by
$H^5(\hat X)$, and by the Serre spectral sequence this is the middle cohomology of
\(
  \label{eq corr}
  H^1(X) \stackrel{(0,d_4)}{\longrightarrow} H^2(X)\oplus H^5(X)
    \stackrel{(d_4,0)}{\longrightarrow}
  H^6(X)
  \,,
\)
where $d_4 = [G_4]\cup(-)$ is the cup product with the degree-4 class of the C-field. For torsion C-fields this
vanishes rationally and hence one arrives at the conclusion that the M-theory super Lie algebra extension is by brane charges
in $H^2(X) \oplus H^5(X)$, agreeing with the result of the argument in \cite{SorokinTownsend97}.
For non-torsion C-fields or else when considered not just rationally, then there are corrections to this
statement by the kernel and cokernel of $d_4$. Notice that these corrections are directly analogous to the
correction by kernel and cokernel of a $d_3$ differential in an Atiyah-Hiruebruch spectral sequence, which appear when refining D-brane charges
from ordinary cohomology to twisted K-theory \cite[(3.2), (3.6)]{MaldacenaMooreSeiberg}.
That 5-brane charges should be in a degree-4 twisted cohomology theory this way has been suggested earlier in
\cite[section 8]{Sati10} and has been discussed further in \cite{FSS15}.



\end{document}